\documentstyle[12pt,titlepage]{article}
\newcommand{\hmn}{\mbox{$\vert H_{-\frac{1}{2}-1}\vert^2$}}

\newcommand{\ho}{\mbox{$\vert H_{-\frac{1}{2}0}\vert^2$}}
\newcommand{\hpl}{\mbox{$\vert H_{\frac{1}{2}1}\vert^2$}}
\newcommand{\hoi}{\mbox{$\vert H_{\frac{1}{2}0}\vert^2$}}

\newcommand{\mbos}{\mbox{{\scriptsize off-shell}}}
\newcommand{\mlab}{\mbox{{\scriptsize lab}}}

\newcommand{\zmin}{z_{\mbox{{\tiny min}}}}
\newcommand{\ra}{\rightarrow}
\newcommand{\lc}{\Lambda_{c}^{+}}

\begin{document}
\begin{titlepage}
\title{On the Determination of the $b \ra c$
  \\Handedness Using Nonleptonic $\Lambda_{c}$-Decays}
\author{B. K\"onig,
J.G. K\"orner\thanks{Supported in part by the BMFT, FRG under contract
 06MZ730}\\
Institut f\"ur Physik,\\
Johannes Gutenberg-Universit\"at,\\
D-55099 Mainz,
Germany \\[0.5cm]
 M. Kr\"amer\thanks{Also at Johannes Gutenberg-Universit\"at Mainz,
 Germany} \thanks{e-mail address: T00KRM@DHHDESY3.DESY.DE}\\
Deutsches Elektronen-Synchrotron DESY\\
D-22603 Hamburg, Germany}
\date{}
\begin{abstract}
We consider possibilities to determine the handedness of $b \ra c$
current transitions using semileptonic baryonic
$\Lambda_{b} \rightarrow \Lambda_{c}$  transitions. We propose
to analyze the longitudinal polarization of the daughter baryon
$\Lambda_c$ by using momentum-spin correlation measurements in the form
of forward-backward (FB) asymmetry measures involving its nonleptonic
decay products.
We use an explicit form factor model to determine the longitudinal
polarization
of the $\Lambda_c$ in the semileptonic decay $\Lambda_b \ra \Lambda_c + l^-
+ \bar \nu_l$. The mean longitudinal polarization of the $\Lambda_c$ is
negative (positive) for left-chiral (right-chiral) $b \ra c$ current
transitions. The frame dependent longitudinal polarization of the $\Lambda_c$
is large ($\cong 80\%$) in the $\Lambda_b$ rest frame and somewhat smaller
(30\% - 40\%) in the lab frame when the $\Lambda_b$'s are produced on the
$Z_0$ peak. We suggest to use nonleptonic decay modes of the $\Lambda_c$ to
analyze its polarization and thereby to determine the chirality of the $b \ra
c$ transition. Since the $\Lambda_b$'s produced on the
$Z_0$ are expected to be polarized we
discuss issues of the polarization transfer in $\Lambda_b \ra
\Lambda_c$ transitions.
We also investigate the $p_\perp$- and $p$-cut sensitivity of
our predictions for the polarization of the $\Lambda_c$.
\end{abstract}
\end{titlepage}
\maketitle
\newpage
In the Standard Model the charged current transition $b \ra c$ is
predicted to be left-chiral, i.e. the Dirac structure of the transition
is given by $\overline{b}
                  \gamma_\mu(1-\gamma_5)c$. This prediction of the
Standard Model has recently been confirmed by a determination of the
sign of the lepton's forward-backward (FB) asymmetry in the ($l^-
\overline{\nu}_l$) rest system in the semileptonic decay
$\overline{B} \ra D^* + l^- +\overline{\nu}_l$
               \cite{argus,cleo}.\footnote{
For a discussion of theoretical background see \cite{ks}.} In this
analysis one uses the Standard Model left-handedness of the lepton
current as input. However, if one leaves the realms of the
Standard Model, the same FB asymmetry would arise if both quark and
lepton currents were taken to be right-chiral, i.e. if one would
switch from a
   $H_{\mu\nu}(V - A) L^{\mu\nu}(V - A)$ coupling to a
   $H_{\mu\nu}(V + A) L^{\mu\nu}(V + A)$ coupling.\footnote{A viable
model involving a right-handed $W_R$ that is
consistent with all present data has recently been proposed \cite{gw}.}

The FB asymmetry measure alluded to above constitutes a
momentum-momentum correlation measure $< \vec{l}\cdot\vec{p}>$ which
clearly is not a truly parity-violating measure.\footnote{For example,
it is well-known that in $e^+e^-$-annihilation the two photon exchange
contribution also gives rise to nonvanishing FB asymmetries despite
of the fact that QED is parity conserving.} What is needed to
distinguish between the two above
                               options is to define truly parity-violating
spin-momentum correlation measures of the type $<\vec{\sigma}\cdot
\vec{p}>$.

Some such possible parity-violating measures that have been discussed
recently       exploit the fact that bottom quarks produced on the
$Z_0$ resonance acquire a $\cong  $ 94\% negative longitudinal
polarization. In the case that the bottom quark hadronizes into the
$\Lambda_b$ bottom baryon there is a 100\% polarization transfer, at
least in the heavy quark limit \cite{close}.
One can then define spin-momentum correlations w.r.t. the
longitudinal spin direction of the decaying $\Lambda_b$ using the
momenta of the decay products of the $\Lambda_b$. For the
semileptonic decays $\Lambda_b \ra \Lambda_c + l^- + \bar\nu_l$
this has been done using the lepton momentum \cite{close,jap} and the
$\Lambda_c$ momentum \cite{jap,kk}. The sign of these correlations or
                                the
sign of the correspondingly defined FB asymmetries allow one to
differentiate the above two options which remain after the analysis
of the mesonic experiments, \cite{argus,cleo}, i.e. the
   $H_{\mu\nu}(V - A) L^{\mu\nu}(V - A)$ or the
   $H_{\mu\nu}(V + A) L^{\mu\nu}(V + A)$ option.
A drawback of the suggested analysis' is that they
require the reconstruction of the $\Lambda_b$ rest frame which
will be a difficult experimental task.\footnote{There is some hope,
though, that such a reconstruction can be done with the newly
installed vertex detectors in the LEP experiments (A.~Putzer, private
communication).}

Alternatively one can consider the shape of the lepton spectrum
directly in the lab system \cite{sample}. The spin-lepton-momentum
correlation effects referred to above have the effect that the
emitted leptons in the semileptonic decay $\Lambda_b \ra \Lambda_c
+ l^- + \bar\nu_l$ (or $b \ra c + l^- + \bar\nu_l$) tend to
counteralign and align with the polarization of the $b$ for
   $H_{\mu\nu}(V - A) L^{\mu\nu}(V - A)$ and
   $H_{\mu\nu}(V + A) L^{\mu\nu}(V + A)$
interactions, respectively,
leading to harder and softer lepton spectra in the lab
                    system relative to unpolarized decay allowing one
to distinguish between the two options in principle. However, as has been
emphasized in
      \cite{close}, a lack of knowledge of the precise form of the
$b \ra \Lambda_b$ fragmentation function precludes a decision
whether the lepton spectrum is harder or softer than that of
unpolarized decay, in particular since there is no unpolarized decay
sample to compare with.

Another possibility to distinguish between the
                       $H_{\mu\nu}(V - A) L^{\mu\nu}(V - A)$ and
$H_{\mu\nu}(V + A) L^{\mu\nu}(V + A)$
                          options via a parity-violating measure is to
determine the polarization of the lepton in the semileptonic decays
$B \ra D(D^*) + l^- + \overline{\nu}_l$ \cite{waik} or
$\Lambda_b \ra \Lambda_c + l^- + \bar\nu_l$ \cite{kkk}.
This will be a difficult experiment but may be feasible in the not too
distant future for semileptonic decays involving the $\tau$-lepton.

In this letter we propose yet a fourth variant of a
                              truly parity-violating
 spin-momentum correlation measure in $b \ra c$ decays. We propose to
look at the decay cascade $\Lambda_b \ra \Lambda_c (\ra a_1 + a_2
                                                          + \cdots)
+ \l^- +\overline{\nu}_l$ to determine the chirality of $b \ra c$
decays where $\Lambda_c \ra a_1 + a_2 + \cdots$ are nonleptonic
decays of the $\Lambda_c$. The weak nonleptonic decays of the
$\Lambda_c$ serve to analyze the polarization of the $\Lambda_c$
through the correlation of their momenta with the polarization of the
decaying $\Lambda_c$. Ideal in this regard are the nonleptonic decays
$\Lambda_c \ra \Lambda \pi$ and
$\Lambda_c \ra \Sigma \pi$ the analyzing power of which has recently
been
determined
\cite{cleo2,argus2,proc}.
                    As a further analyzing channel we discuss the decay
modes
$\lc \ra p \bar K^{*0}$ and $\lc \ra \Delta^{++} K^-$ which could make up
a large fraction of the dominant decay mode
 $\Lambda_c \ra p  K^- \pi^+$. The analyzing power of these channels has
not yet been determined experimentally but can be estimated using
the theoretical quark model ansatz of \cite{kk2}.

Consider first the semileptonic decay of an unpolarized $\Lambda_b$.
Possible polarization effects due to polarized $\Lambda_b$-decays
average out if one integrates over all possible momentum directions
of the $\Lambda_c$ in the decay $\Lambda_b \ra \Lambda_c + l^- +
\overline{\nu}_l$. Possible $\Lambda_b$ polarization
effects due to incomplete averaging because of
experimental cut biases will be discussed later on. We define helicity
form factors for the $\Lambda_b \ra \Lambda_c$ transition in the
 $\Lambda_b$ rest system by writing
\begin{equation}
H_{\lambda_2\lambda_W} = \langle \Lambda_2; \lambda_2 |V_\mu -
\xi A_\mu|
\Lambda_1; \lambda_1 \rangle
                             \epsilon^{\mu}
                                        (\lambda_W)
\end{equation}
where we have switched to a more generic notation and identify the
labels $b$ and $c$ with 1 and 2, respectively.
We have introduced a chirality parameter $\xi$ which takes the value
$\xi = 1$ and $\xi = -1$ for left-chiral and right-chiral current
transitions, respectively.
$\lambda_i$ and
$\lambda_W$ denote the helicities of the $\Lambda_i$ ($i = 1,2$) and
the off-shell $W$-Boson where $\lambda_1 = \lambda_2 - \lambda_W$
\cite{kk,bkkz}. The longitudinal polarization $P_L$ of the $\Lambda_c$
along the momentum direction of the $\Lambda_c$ in the $\Lambda_b$
{\em rest system} is
given by \cite{kk,bkkz}\footnote{In Ref.\cite{kk} the longitudinal
polarization was denoted by $\alpha$.}
                        (the polarization of the $\Lambda_c$ in the lab
frame will be discussed later on)
\begin{equation}\label{pl}
P_L = \frac{\hpl - \hmn + \hoi - \ho}{\hpl + \hmn + \hoi + \ho} \quad .
\end{equation}
Employing simple helicity arguments $P_L$ is expected to be
negative and positive in most of the phase space region
for left-chiral ($\xi = 1$) and right-chiral ($\xi = -1$) $b\ra c$
transitions, respectively. For the mean value of $P_L$ one finds
\begin{equation}\label{plmean}
\langle P_L\rangle
 = \xi \bigg\{ \!\begin{array}{ll} -0.77 & \mbox{IMF}\; \cite{kkkk}
        \\ -0.81 & \mbox{FQD} \end{array} \quad .
\end{equation}
The two polarization values refer to
 the Heavy Quark Effective Theory (HQET) improved infinite momentum
frame (IMF) model of Ref.\cite{kkkk} and free quark decay (FQD) where we use
$m_b = M_{\Lambda_b} = 5.64$ GeV and $m_c = M_{\Lambda_c} = 2.285$ GeV
in order to get the phase space right (see e.g. \cite{kkkk}).\footnote{The
difference in the two values Eq.(\ref{plmean}) does not imply that $1/m_Q$
effects are large in the IMF model of \cite{kkkk}. The difference is
mainly due to
form factor
effects which enhance the high $q^2$-region in form factor models where the
polarization is smallest.}

The longitudinal polarization of the $\Lambda_c$ can be probed by
looking at the angular distribution of its subsequent nonleptonic
decays. Ideal in this regard are the nonleptonic modes
$\Lambda_c \ra \Lambda\pi$ and $\Lambda_c \ra \Sigma\pi$
since the analyzing power of these decays has recently been
determined. For $\Lambda_c \ra \Lambda\pi$ one has
\begin{equation}\label{alp1}
\alpha_{\Lambda_c\ra\Lambda\pi}
 = \bigg\{ \!\begin{array}{lc} -1.0 {+ 0.4 \atop - 0.0}& \cite{cleo2}
        \\ -0.96 \pm 0.42 & \cite{argus2} \end{array} \quad .
\end{equation}
For $\Lambda_c \ra \Sigma\pi$ we quote the preliminary value \cite{proc}
\begin{equation}\label{alp}
\hphantom{deca }
\alpha_{\Lambda_c\ra\Sigma\pi} = - 0.43 \pm 0.23 \pm 0.20 \quad .
\end{equation}
The decay distribution of the $\Lambda$ or $\Sigma$ in the
$\Lambda_c$ rest frame
reads \cite{kk,bkkz}
\begin{equation}\label{decaydist}
W(\Theta_\Lambda) = 1 +     P_L \alpha_{\Lambda_c}\cos\Theta
\end{equation}
where the polar angle $\Theta$ is measured w.r.t. the original
 flight direction of the $\Lambda_c$ and $\alpha_{\Lambda_c}$ stands for
either of the asymmetry parameters in (\ref{alp1},\ref{alp}).
 Correspondingly one can define a
forward-backward (FB) asymmetry by averaging over the daughter baryons
in the
respective forward (F) $(0^\circ \le\Theta < 90^\circ)$ and backward (B)
$(90^\circ \le \Theta < 180^\circ)$ hemispheres to obtain
\begin{equation}\label{fb}
A_{FB} = \frac{1}{2}    P_L \alpha_{\Lambda_c} \quad .
\end{equation}
Judging from the large numerical values of the mean of
                                           $P_L$ Eq.(\ref{plmean}) and
 of the asymmetry parameters
$\alpha_{\Lambda_c}$ Eqs.(\ref{alp1},\ref{alp})
                                   a measurement of the sign of $A_{FB}$
within reasonable errors should allow one to conclude for the sign of
$\xi$ and therefore for the chirality of the $b \ra c$ transition with
a good certainty.

Next we turn to the decay mode $\Lambda_c \ra p K^- \pi^+$. This is
the darling channel for experimentalists as it is easy to identify
experimentally. According to \cite{argcle} its branching ratio is
approximately five times bigger than $\Lambda_c \ra \Lambda\pi$. Note
also that this decay mode has been used to reconstruct the
$\Lambda_c$ in semileptonic $\Lambda_b$ decays produced on the $Z_0$
\cite{aleph}.
However, nothing is known experimentally about the analyzing power of
this channel. We therefore have to turn to some theoretical input.
One may either concentrate on the resonant substructures $\Lambda_c
\ra p \bar K^{*0} $ and $\Lambda_c \ra \Delta^{++} K^-$
present in $\Lambda_c \ra p K^{-} \pi^{+}$ or
         treat the decay in a resonance approximation in that one
        assumes
that the decay is dominated by the channels $\Lambda_c \ra p
\bar      K^{*0}$ and $\Lambda_c \ra \Delta^{++} K^-$.
The present
experimental evidence for the viability of such a resonance
approximation is somewhat inconclusive. The Mark II collaboration
             \cite{weiss}
quotes relative branching ratios of $(18\pm 10)\%$ and $(17\pm 7)\%$
for $\Lambda \ra p \bar      K^{*0}$ and $\Lambda_c \ra \Delta^{++}
K^-$, resp., relative to $\Lambda_c \ra p K^+ \pi^-$,    the R415
collaboration \cite{basile}
                      quotes $(42\pm 24)\%$ and $(40\pm 17)\%$, resp.,
for the same two relative branching ratios and, more recently, the
ACCMOR collaboration \cite{accmor} quotes
                 ($35{+0.06 \atop -0.07}\pm0.03)\%$
 and $(12{+0.04\atop -0.05}\pm 0.05)\%$, resp. One can only hope that
future experiments can clarify the situation. At any rate, the channel
$\Lambda_c \ra p \bar K^{*0}$ can be expected to have a substantial
branching ratio.

For the decay mode $\Lambda_{c}^{\uparrow} \ra p \bar      K^{*0}$
one can write down a polar decay distribution in complete analogy to
Eq.(\ref{decaydist}). In the $\Lambda_c$ rest frame one has
\begin{equation}\label{decaydist2}
W(\Theta_p) = 1 +     P_L \alpha_p \cos\Theta_p
\end{equation}
where $\Theta_p$ is the polar angle of the proton relative to the
original direction of flight of the $\Lambda_c$.
The asymmetry parameter $\alpha_p$ is given by
\begin{equation}\label{alphap}
\alpha_p = \frac{- \hpl + \hmn + \hoi - \ho}{\hpl + \hmn + \hoi + \ho}
\end{equation}
and the $H_{\lambda_p
                  \lambda_{K^*}}$ are helicity amplitudes defined by
 (see e.g. \cite{kk2})
\begin{equation}
H_{\lambda_p \lambda_{K^{*}}} = \langle p, \lambda_p; \bar      K^{*0}
,\lambda_{K^*} | {\cal H}_{n.l.} | \Lambda_c, \lambda_{\Lambda_c}\rangle
\end{equation}
with $\lambda_p - \lambda_{K^*} = \lambda_{\Lambda_c}$. We mention that
the decay distribution Eq.(\ref{decaydist2})
                                   and the asymmetry parameter
$\alpha_p$ (\ref{alphap}) can be directly transcribed from the
corresponding decay distribution for $(1/2^{+})^{\uparrow}\ra (1/2^{+})
+ W_{\mbos} $ written down in \cite{kk,bkkz}.

Analogous to Eq.(\ref{fb}) one can then define a forward-backward
asymmetry averaging over protons in the forward (F) ($0^\circ \le \Theta
< 90^\circ$) and backward (B) ($90^\circ \le \Theta < 180^\circ$)
                                                            hemispheres,
where F and B are defined relative to the flight direction of the
$\Lambda_c$. One obtains
\begin{equation}
A_{FB} = \frac{1}{2}     P_L \alpha_p  \quad .
\end{equation}

The asymmetry parameter $\alpha_p$ can be calculated using the quark
model approach of Ref.\cite{kk2}. The relevant quark line diagrams are
drawn in Fig.~1. For the decay $\Lambda_c \ra p \bar      K^{*0}$
there is a factorizing contribution (2a) and a $W$-exchange
contribution (2b). The relative amplitude of the two contributions
has been determined in \cite{kk2} through a fit to the available data
on nonleptonic $\Lambda_c$ decays whereas the factorizing contribution
can be calculated for particular wave function models. Using the
results of \cite{kk2} one finds
\begin{eqnarray}\label{ha}
H_{\frac{1}{2}1} &=& (2.14 - 0.40) \times 10^{-6} \nonumber \\
H_{-\frac{1}{2}-1} &=& (-3.24 - 1.58) \times 10^{-6} \nonumber \\
H_{\frac{1}{2}0} &=& (-1.46 - 1.68) \times 10^{-6} \nonumber \\
H_{-\frac{1}{2}0} &=& (4.26 - 2.51) \times 10^{-6}
\end{eqnarray}
where the two numbers in the round brackets refer to the contributions
of diagrams (2a) and (2b), respectively. The contributions of the
factorizing contribution                                      (2a) and
the $W$-exchange contribution
(2b) are constructive for the helicity amplitudes $H_{-\frac{1}{2}-1}$
and $H_{\frac{1}{2}0}$ and destructive for the helicity amplitudes
$H_{\frac{1}{2}1}$ and $H_{-\frac{1}{2}0}$. It is therefore clear that
one will have a negative asymmetry value and thereby a negative value
for $A_{FB}$ for the left-chiral $b \ra c$ currents. Numerically one
obtains
\begin{equation}\label{12}
\alpha_p = 0.69
\end{equation}
using the model values (\ref{ha}). Note, though, that the predicted
value Eq.(\ref{12}) is quite sensitive to the relative weight and
sign of the contributions written down in (\ref{ha}) (factorizing and
nonfactorizing) and is thereby subject to some theoretical
uncertainty.

Concerning
the channel $\Lambda_c \ra \Delta^{++}K^-$ one notes that this decay
is contributed to only by
the $W$-exchange diagram as drawn in Fig.~1c. One has the two
helicity amplitudes $H_{\lambda_\Delta\lambda_\pi}$ with $\lambda_\Delta =
\pm 1/2$.
Looking at the helicity
configurations of the quark diagrams one finds $H_{\frac{1}{2}0}
= H_{-\frac{1}{2}0}$ because of the symmetric nature of the $\Delta^{++}
$ quark model wave function. Thus one finds that the decay
$\Lambda_c \ra \Delta^{++} K^-$ is a purely parity conserving
$p$-wave transition \cite{kk2}. Correspondingly the asymmetry parameter
in this decay is zero.

                 If one considers the sum
of the two above subchannels one finds a
diluted asymmetry value
for the asymmetry of the proton in the decay $\Lambda_c \ra p K^{*0}
+ \Delta^{++} K^-$.
One then has
\begin{equation}
\alpha_p = 0.37 \mbox{-} 0.46
\end{equation}
where the first and second value refer to a 88\% and 50\%
ratio of the $\Lambda_c \ra
\Delta^{++} K^{-}$ and  $\Lambda_c \ra p \bar K^{*0}$ rates.

Summarizing our results for the two subchannels of
 $\Lambda_c \ra p K^- \pi^+$ considered by us
 we find that the proton is preferentially emitted backward (forward)
 for a left(right)-chiral $b \ra c$ transition. The analyzing power
of this nonleptonic decay mode is large in particular if one selects
the $\Lambda_c \ra p \bar      K^{*0}$ band.

Let us now return to the question of polarization transfer from a
polarized $\Lambda_b$ with longitudinal
                                        polarization $P$ ($-1 \le P
\le 1$) to a polarized $\Lambda_c$ with longitudinal polarization
$P_L$ ($ -1 \le P_L \le 1$).
To this end we write down the unnormalized
density matrix elements of the $\Lambda_c$ in the $\Lambda_b$ rest
system \cite{kk}:
\begin{eqnarray}
\rho_{\frac{1}{2}\frac{1}{2}}(\cos\Theta_{\Lambda_c})
                             &=&\hpl (1 - P\cos\Theta_{\Lambda_c})
+ \hoi (1 + P\cos\Theta_{\Lambda_c}) \nonumber\\
\rho_{-\frac{1}{2} -\frac{1}{2}}(\cos\Theta_{\Lambda_c})
                                &=&\hmn (1 + P\cos\Theta_{\Lambda_c})
+ \ho (1 - P\cos\Theta_{\Lambda_c})
\end{eqnarray}
where $\Theta_{\Lambda_c}$ is the polar angle of the $\Lambda_c$
relative to the original flight direction of the $\Lambda_b$ in the
$\Lambda_b$
rest frame. The
$\cos\Theta_{\Lambda_c}$ dependence of the longitudinal polarization
$P_L$ of the $\Lambda_c$ can then be calculated from
\begin{equation}\label{pl2}
P_L(\cos\Theta_{\Lambda_c}) =
\frac{\rho_{\frac{1}{2}\frac{1}{2}}(\cos\Theta_{\Lambda_c}) -
\rho_{-\frac{1}{2}-\frac{1}{2}}(\cos\Theta_{\Lambda_c})}
{\rho_{\frac{1}{2}\frac{1}{2}}(\cos\Theta_{\Lambda_c}) +
\rho_{-\frac{1}{2}-\frac{1}{2}}(\cos\Theta_{\Lambda_c})}  \quad .
\end{equation}

In Fig.~2 we show the $\cos\Theta_{\Lambda_c}$-dependence of
$<P_L>$ of $\Lambda_c$ again for
the HQET improved IMF model of \cite{kkkk} and the FQD model.
For definiteness we have taken $P = -0.94$. This refers to the case of
$\Lambda_b$'s produced on the $Z_0$. As mentioned in the Introduction $b$
quarks produced on the $Z_0$ are expected to be negatively polarized with a
94\% degree of polarization. Here we assume that the polarization transfer
in the fragmentation $b \ra \Lambda_b$ is 100\%, as predicted in the heavy
quark limit \cite{close}. For smaller values of $P$ the asymmetry in the
polarization transfer plot Fig.~2 would be reduced.
At
$90^\circ$ there clearly is no polarization transfer and one recovers
the values of Eq.(\ref{plmean}).
The polarization transfer in Fig.~2 has been calculated for left-chiral
($\xi = 1$) $b \ra c$ transitions. The right-chiral case ($\xi = - 1$) is
obtained from Fig.~2 by the replacement $P_L \ra - P_L$ and
$\Theta_{\Lambda_c} \ra \pi - \Theta_{\Lambda_c}$, i.e. reflections on both
zero axis'.
 As emphasized
above the dependence of $P_L$ on $P$ drops out when one
integrates over $\cos\Theta_{\Lambda_c}$.

What has been said up to now requires the reconstruction of the
$\Lambda_b$ rest system. This will not be an easy task for the
energetic $\Lambda_b$ bottom baryons produced on the $Z_0$ where the
analysis suggested in this paper is most likely to be done first.
There is some hope, though, that such a reconstruction can be done
with the newly installed vertex detectors in the CERN detectors,
as mentioned before.
                                    Nevertheless we shall in the
following discuss the more realistic situation present in the LEP
environment of energetic longitudinally polarized $\Lambda_b$'s
whose rest frames cannot be reconstructed. The polarization of
the $\Lambda_c$'s in the semileptonic decays takes a more
complicated form in the laboratory frame than in the $\Lambda_b$
rest frame as given by Eq.(\ref{pl}) and (\ref{pl2}). In particular
negatively polarized $\Lambda_c$'s emerging backward in the
$\Lambda_b$ rest frame will turn into positively polarized
$\Lambda_c$'s in the lab frame because of the momentum reversal
due to the requisite Lorentz boost. Also, because of experimental
cuts and/or biases the $\Lambda_c$'s polarization dependence on the
polarization of the $\Lambda_b$ may no longer average out, i.e. one
has to address the question of polarization transfer under realistic
experimental conditions.

In order to study all these issues we have written a Monte Carlo program that
generates semileptonic decay events of polarized $\Lambda_b$ into polarized
$\Lambda_c$. It is then a simple matter
to adapt our calculation to the experimental conditions present in
the LEP environment including longitudinal and transverse lepton momentum
cuts.

In Fig.~3 the dependence of $<P_L>$ on the energy of the $\Lambda_b$
in the lab frame is shown for the FQD model with $m_b = m_{\Lambda_b}
= 5.64$~GeV and $m_c = m_{\Lambda_c} = 2.285$~GeV where $E_{\Lambda_b} =
z \cdot M_Z/2$. At $\zmin =
2m_{\Lambda_b}/M_Z$ corresponding to a $\Lambda_b$ being produced at
rest we
have $<P_L> = -0.81$ as given in Eq.(3). For $\zmin < z
\stackrel{\textstyle<}{\sim}
0.3$ the mean polarization
$<P_L>$ quickly increases and shows almost no $z$-dependence for $z
\stackrel{\textstyle>}{\sim}
0.3$. The reason that the mean polarization of the $\Lambda_c$ saturates so
fast is clear: the average energy released in $\Lambda_b \ra \Lambda_c +
l^- + \bar\nu_l$ is quite small on the scale of the $Z_0$-mass. In particular
the sign of the longitudinal polarization does not change
over the whole $z$-range. The same behaviour is true
 for the IMF quark model calculation of \cite{kkkk}.

It is obvious from Fig.~3 that our results are practically not affected by
the details of
fragmentation: the fragmentation function $b \ra \Lambda_b$
is expected to be strongly
peaked in the high $z$ region where the saturation of $<P_L>$ has set in.
This is born out by the so called Peterson fragmentation function
\cite{peter}.
Further we conclude that our predictions for $<P_L>_{\mlab}$ will only
be marginally
affected by the folding in of any realistic fragmentation function.

The last point we want to discuss is the cut dependence of our predictions for
the $\Lambda_c$'s polarization. The cut dependence comes in because of
experimental trigger requirements: one triggers on high $p_\perp$ and
high $p$
leptons in order to select on semileptonic $\Lambda_b$ decays
\cite{aleph,opal}. Again we use a polarization of $P = - 0.94$ for the
$b$-quark and for $\Lambda_b$.
As can be judged
from the numbers in Table~1 the effects of such cuts have little effect on our
prediction for the polarization of the $\Lambda_c$ in the lab frame.
There is a
small effect in that the cuts tend to enhance the longitudinal polarization
in the lab frame

\begin{table}\label{pl3}
\begin{tabular}{|l|cc|}
\hline
$<P_L>$ & FQD model & quark model \cite{kkkk} \\
\hline
\hline
$\Lambda_b$ rest frame & $-0.81$ & $-0.77$ \\
lab frame; no cuts & $-0.36$ & $-0.26$ \\
lab frame; cut on $p_\perp$ & $-0.41$ & $-0.32$ \\
lab frame; cut on $p_\perp$ and $p$ & $-0.40$ & $-0.31$ \\
\hline
\end{tabular}
\parbox{11.5cm}{
\caption{Values for the mean longitudinal polarization $<P_L>$
of the $\Lambda_c$ in the $\Lambda_b$ rest frame and in the lab frame from
$Z_0$-decays with and without cuts. The energy of the $\Lambda_b$
in the lab frame is taken to be 40~GeV corresponding to a mean value of
$<z> \approx 0.88$ (cf. [23]). We use $p_{\perp}^{cut}
= 1$~GeV and $p^{cut} = 3$~GeV [18, 22].}}
\end{table}

Table~1 summarizes our results on the calculation of $<P_L>$. We find
a large longitudinal polarization of the $\Lambda_c$ in the
$\Lambda_b$ rest frame leading to  large forward-backward asymmetries
in subsequent nonleptonic decays of the $\Lambda_c$.
The absolute value of the longitudinal polarization (and thereby the
forward-backward asymmetry) is reduced by about a factor of two
when the analysis has to be performed in the LEP lab frame. Our
predictions are practically not affected by fragmentation and possible
experimental cuts.

In summary we have used an explicit form factor model and the free quark decay
model to determine the longitudinal polarization of the $\Lambda_c$ in the
semileptonic decays $\Lambda_b \ra \Lambda_c + l^- + \bar\nu_l$. The mean
longitudinal polarization of the $\Lambda_c$ is negative (positive) for the
left-chiral (right-chiral) $b\ra c$ current transitions. The mean longitudinal
polarization of the $\Lambda_c$ turns out to be large ($\cong 80\%$) in the
$\Lambda_b$ rest frame and somewhat smaller (30\% - 40\%) in the lab
frame when
$\Lambda_b$'s are produced on the $Z_0$-peak. We have suggested to use
nonleptonic decay modes of the $\Lambda_c$ to analyse its polarization. Most
useful in this regard are the decay modes $\Lambda_c \ra \Lambda \pi$ and
$\Lambda_c \ra \Sigma\pi$ since the decay asymmetry parameters in these modes
have recently been measured. We have also discussed the modes $\Lambda_c \ra p
\bar K^{*0}$ and $\Lambda_c \ra \Delta^{++}K^-$ for which we have provided
theoretical model dependent decay asymmetry parameters. We believe that the
issue whether the $b \ra c$ transitions are left- or right-chiral can be
settled in the near future using the analysis suggested in this paper.

\noindent
Acknowledgement: Part of this work was done while J.G.K. was a visitor at the
DESY theory group. He would like to thank W.~Buchm\"uller for the hospitality
and the DESY directorate for support.

\newpage

\noindent
Figure Captions

\noindent
\newcounter{fig}
\begin{list}{\bf Fig. \arabic{fig}:}{\usecounter{fig}
  \labelwidth1.6cm \leftmargin2.5cm \labelsep0.4cm \rightmargin1cm
  \parsep0.5ex plus0.2ex minus0.1ex \itemsep0ex plus0.2ex }
\item Flavour diagrams contributing to two-body nonleptonic decays of the
      $\Lambda_c$. For illustrative purposes we have labelled the
      flavour diagrams according to the decay $\Lambda_c \ra \Lambda \pi$.
\item Polarization transfer from a 94\% (negatively) longitudinally polarized
      $\Lambda_c$ in semileptonic decays
      $\Lambda_b \ra \Lambda_c + l^- + \bar\nu_l$
      as a function of the angle $\Theta_{\Lambda_c}$ between the
      $\Lambda_c$ and the $\Lambda_b$.
\item Mean longitudinal polarization of lab frame $\Lambda_c$'s from
      $\Lambda_b$'s produced on the $Z_0$ as a function of the $\Lambda_b$'s
      fractional energy.
\end{list}
\end{document}